
%
%
%
\def\unredoffs{} \def\redoffs{\voffset=-.31truein\hoffset=-.59truein}
\def\speclscape{\special{ps: landscape}}
%
%
%
%
\newbox\leftpage \newdimen\fullhsize \newdimen\hstitle \newdimen\hsbody
\tolerance=1000\hfuzz=2pt
\catcode`\@=11 
\def\bigans{b }\def\answ{b }
%
\ifx\answ\bigans\message{(This will come out unreduced.}
\magnification=1200\unredoffs\baselineskip=8pt plus 2pt minus 1pt
\hsbody=\hsize \hstitle=\hsize 
\else\message{(This will be reduced.} \let\l@r=L
\magnification=1000\baselineskip=16pt plus 2pt minus 1pt \vsize=7truein
\redoffs \hstitle=8truein\hsbody=4.75truein\fullhsize=10truein\hsize=\hsbody
\output={\ifnum\pageno=0 
  \shipout\vbox{\speclscape{\hsize\fullhsize\makeheadline}
    \hbox to \fullhsize{\hfill\pagebody\hfill}}\advancepageno
  \else
  \almostshipout{\leftline{\vbox{\pagebody\makefootline}}}\advancepageno
  \fi}
\def\almostshipout#1{\if L\l@r \count1=1 \message{[\the\count0.\the\count1]}
      \global\setbox\leftpage=#1 \global\let\l@r=R
 \else \count1=2
  \shipout\vbox{\speclscape{\hsize\fullhsize\makeheadline}
      \hbox to\fullhsize{\box\leftpage\hfil#1}}  \global\let\l@r=L\fi}
\fi
%
\newcount\yearltd\yearltd=\year\advance\yearltd by -1900

%

%

\def\draftmode{\message{ DRAFTMODE }\def\draftdate{{\rm preliminary draft:
\number\month/\number\day/\number\yearltd\ \ \hourmin}}%
\headline={\hfil\draftdate}\writelabels\baselineskip=20pt plus 2pt minus 2pt
 {\count255=\time\divide\count255 by 60 \xdef\hourmin{\number\count255}
  \multiply\count255 by-60\advance\count255 by\time
  \xdef\hourmin{\hourmin:\ifnum\count255<10 0\fi\the\count255}}}
\def\nolabels{\def\wrlabeL##1{}\def\eqlabeL##1{}\def\reflabeL##1{}}
\def\writelabels{\def\wrlabeL##1{\leavevmode\vadjust{\rlap{\smash%
{\line{{\escapechar=` \hfill\rlap{\sevenrm\hskip.03in\string##1}}}}}}}%
\def\eqlabeL##1{{\escapechar-1\rlap{\sevenrm\hskip.05in\string##1}}}%
\def\reflabeL##1{\noexpand\llap{\noexpand\sevenrm\string\string\string##1}}}
\nolabels
%
\global\newcount\secno \global\secno=0
\global\newcount\meqno \global\meqno=1
\def\newsec#1{\global\advance\secno by1\message{(\the\secno. #1)}
\global\subsecno=0\eqnres@t\noindent{\bf\the\secno. #1}
\writetoca{{\secsym} {#1}}\par\nobreak\medskip\nobreak}
\def\eqnres@t{\xdef\secsym{\the\secno.}\global\meqno=1\bigbreak\bigskip}
\def\sequentialequations{\def\eqnres@t{\bigbreak}}\xdef\secsym{}
\global\newcount\subsecno \global\subsecno=0
\def\subsec#1{\global\advance\subsecno by1\message{(\secsym\the\subsecno. #1)}
\ifnum\lastpenalty>9000\else\bigbreak\fi
\noindent{\it\secsym\the\subsecno. #1}\writetoca{\string\quad
{\secsym\the\subsecno.} {#1}}\par\nobreak\medskip\nobreak}
\def\appendix#1#2{\global\meqno=1\global\subsecno=0\xdef\secsym{\hbox{#1.}}
\bigbreak\bigskip\noindent{\bf Appendix #1. #2}\message{(#1. #2)}
\writetoca{Appendix {#1.} {#2}}\par\nobreak\medskip\nobreak}
%
%
\def\eqnn#1{\xdef #1{(\secsym\the\meqno)}\writedef{#1\leftbracket#1}%
\global\advance\meqno by1\wrlabeL#1}
\def\eqna#1{\xdef #1##1{\hbox{$(\secsym\the\meqno##1)$}}
\writedef{#1\numbersign1\leftbracket#1{\numbersign1}}%
\global\advance\meqno by1\wrlabeL{#1$\{\}$}}
\def\eqn#1#2{\xdef #1{(\secsym\the\meqno)}\writedef{#1\leftbracket#1}%
\global\advance\meqno by1$$#2\eqno#1\eqlabeL#1$$}
%
\newskip\footskip\footskip14pt plus 1pt minus 1pt 
\def\footnotefont{\ninepoint}\def\f@t#1{\footnotefont #1\@foot}
\def\f@@t{\baselineskip\footskip\bgroup\footnotefont\aftergroup\@foot\let\next}
\setbox\strutbox=\hbox{\vrule height9.5pt depth4.5pt width0pt}
\global\newcount\ftno \global\ftno=0
\def\foot{\global\advance\ftno by1\footnote{$^{\the\ftno}$}}
%
\newwrite\ftfile
\def\footend{\def\foot{\global\advance\ftno by1\chardef\wfile=\ftfile
$^{\the\ftno}$\ifnum\ftno=1\immediate\openout\ftfile=foots.tmp\fi%
\immediate\write\ftfile{\noexpand\smallskip%
\noexpand\item{f\the\ftno:\ }\pctsign}\findarg}%
\def\footatend{\vfill\eject\immediate\closeout\ftfile{\parindent=20pt
\centerline{\bf Footnotes}\nobreak\bigskip\input foots.tmp }}}
\def\footatend{}
%
%
\global\newcount\refno \global\refno=1
\newwrite\rfile
\def\ref{$^{\the\refno}$\nref}
\def\tref{[\the\refno]\nref}
\def\nref#1{\xdef#1{\the\refno}\writedef{#1\leftbracket#1}%
\ifnum\refno=1\immediate\openout\rfile=refs.tmp\fi
\global\advance\refno by1\chardef\wfile=\rfile\immediate
\write\rfile{\noexpand\item{#1.\ }\reflabeL{#1\hskip.31in}\pctsign}\findarg}
\def\findarg#1#{\begingroup\obeylines\newlinechar=`\^^M\pass@rg}
{\obeylines\gdef\pass@rg#1{\writ@line\relax #1^^M\hbox{}^^M}%
\gdef\writ@line#1^^M{\expandafter\toks0\expandafter{\striprel@x #1}%
\edef\next{\the\toks0}\ifx\next\em@rk\let\next=\endgroup\else\ifx\next\empty%
\else\immediate\write\wfile{\the\toks0}\fi\let\next=\writ@line\fi\next\relax}}
\def\striprel@x#1{} \def\em@rk{\hbox{}}
\def\lref{\begingroup\obeylines\lr@f}
\def\lr@f#1#2{\gdef#1{\ref#1{#2}}\endgroup\unskip}

\def\addref#1{\immediate\write\rfile{\noexpand\item{}#1}} 
\def\startrefs#1{\immediate\openout\rfile=refs.tmp\refno=#1}
\def\xref{\expandafter\xr@f}\def\xr@f[#1]{#1}
\def\refs#1{\count255=1[\r@fs #1{\hbox{}}]}
\def\r@fs#1{\ifx\und@fined#1\message{reflabel \string#1 is undefined.}%
\nref#1{need to supply reference \string#1.}\fi%
\vphantom{\hphantom{#1}}\edef\next{#1}\ifx\next\em@rk\def\next{}%
\else\ifx\next#1\ifodd\count255\relax\xref#1\count255=0\fi%
\else#1\count255=1\fi\let\next=\r@fs\fi\next}
%

%
\newwrite\ffile\global\newcount\figno \global\figno=1
\def\fig{fig.~\the\figno\nfig}
\def\nfig#1{\xdef#1{fig.~\the\figno}%
\writedef{#1\leftbracket fig.\noexpand~\the\figno}%
\ifnum\figno=1\immediate\openout\ffile=figs.tmp\fi\chardef\wfile=\ffile%
\immediate\write\ffile{\noexpand\medskip\noexpand\item{Fig.\ \the\figno. }
\reflabeL{#1\hskip.55in}\pctsign}\global\advance\figno by1\findarg}
\def\xfig{\expandafter\xf@g}\def\xf@g fig.\penalty\@M\ {}
\def\figs#1{figs.~\f@gs #1{\hbox{}}}
\def\f@gs#1{\edef\next{#1}\ifx\next\em@rk\def\next{}\else
\ifx\next#1\xfig #1\else#1\fi\let\next=\f@gs\fi\next}
\newwrite\lfile
{\escapechar-1\xdef\pctsign{\string\%}\xdef\leftbracket{\string\{}
\xdef\rightbracket{\string\}}\xdef\numbersign{\string\#}}

\def\writestop{\def\writestoppt{\immediate\write\lfile{\string\pageno%
\the\pageno\string\startrefs\leftbracket\the\refno\rightbracket%
\string\def\string\secsym\leftbracket\secsym\rightbracket%
\string\secno\the\secno\string\meqno\the\meqno}\immediate\closeout\lfile}}
\def\writestoppt{}\def\writedef#1{}
\def\seclab#1{\xdef #1{\the\secno}\writedef{#1\leftbracket#1}\wrlabeL{#1=#1}}
\def\subseclab#1{\xdef #1{\secsym\the\subsecno}%
\writedef{#1\leftbracket#1}\wrlabeL{#1=#1}}
\newwrite\tfile \def\writetoca#1{}
\def\leaderfill{\leaders\hbox to 1em{\hss.\hss}\hfill}
\def\writetoc{\immediate\openout\tfile=toc.tmp
   \def\writetoca##1{{\edef\next{\write\tfile{\noindent ##1
   \string\leaderfill {\noexpand\number\pageno} \par}}\next}}}

%
\catcode`\@=12 
%
\edef\tfontsize{\ifx\answ\bigans scaled\magstep3\else scaled\magstep4\fi}
 \tfontsize  \tfontsize
 \tfontsize \font\titlei=cmmi10 \tfontsize
\font\titleis=cmmi7 \tfontsize \font\titleiss=cmmi5 \tfontsize
\font\titlesy=cmsy10 \tfontsize \font\titlesys=cmsy7 \tfontsize
\font\titlesyss=cmsy5 \tfontsize  \tfontsize
\skewchar\titlei='177 \skewchar\titleis='177 \skewchar\titleiss='177
\skewchar\titlesy='60 \skewchar\titlesys='60 \skewchar\titlesyss='60
 \ifx\answ\bigans\else scaled\magstep1\fi
\ifx\answ\bigans\def\abstractfont{\ninerm}\else
\font\abssl=cmsl10 scaled \magstep1
\font\absrm=cmr10 scaled\magstep1 \font\absrms=cmr7 scaled\magstep1
\font\absrmss=cmr5 scaled\magstep1 \font\absi=cmmi10 scaled\magstep1
\font\absis=cmmi7 scaled\magstep1 \font\absiss=cmmi5 scaled\magstep1
\font\abssy=cmsy10 scaled\magstep1 \font\abssys=cmsy7 scaled\magstep1
\font\abssyss=cmsy5 scaled\magstep1 \font\absbf=cmbx10 scaled\magstep1
\skewchar\absi='177 \skewchar\absis='177 \skewchar\absiss='177
\skewchar\abssy='60 \skewchar\abssys='60 \skewchar\abssyss='60
\def\abstractfont{\def\rm{\fam0\absrm}
\textfont0=\absrm \scriptfont0=\absrms \scriptscriptfont0=\absrmss
\textfont1=\absi \scriptfont1=\absis \scriptscriptfont1=\absiss
\textfont2=\abssy \scriptfont2=\abssys \scriptscriptfont2=\abssyss
\textfont\itfam=\bigit \def\it{\fam\itfam\bigit}\def\footnotefont{\tenpoint}%
\textfont\slfam=\abssl \def\sl{\fam\slfam\abssl}%
\textfont\bffam=\absbf \def\bf{\fam\bffam\absbf}\rm}\fi
\def\tenpoint{\def\rm{\fam0\tenrm}
\textfont0=\tenrm \scriptfont0=\sevenrm \scriptscriptfont0=\fiverm
\textfont1=\teni  \scriptfont1=\seveni  \scriptscriptfont1=\fivei
\textfont2=\tensy \scriptfont2=\sevensy \scriptscriptfont2=\fivesy
\textfont\itfam=\tenit \def\it{\fam\itfam\tenit}\def\footnotefont{\ninepoint}%
\textfont\bffam=\tenbf \def\bf{\fam\bffam\tenbf}\def\sl{\fam\slfam\tensl}\rm}
\font\ninerm=cmr9 \font\sixrm=cmr6 \font\ninei=cmmi9 \font\sixi=cmmi6
\font\ninesy=cmsy9 \font\sixsy=cmsy6 \font\ninebf=cmbx9
\font\nineit=cmti9 \font\ninesl=cmsl9 \skewchar\ninei='177
\skewchar\sixi='177 \skewchar\ninesy='60 \skewchar\sixsy='60
\def\ninepoint{\def\rm{\fam0\ninerm}
\textfont0=\ninerm \scriptfont0=\sixrm \scriptscriptfont0=\fiverm
\textfont1=\ninei \scriptfont1=\sixi \scriptscriptfont1=\fivei
\textfont2=\ninesy \scriptfont2=\sixsy \scriptscriptfont2=\fivesy
\textfont\itfam=\ninei \def\it{\fam\itfam\nineit}\def\sl{\fam\slfam\ninesl}%
\textfont\bffam=\ninebf \def\bf{\fam\bffam\ninebf}\rm}
%
%

\hyphenation{anom-aly anom-alies coun-ter-term coun-ter-terms}
\def\inv{^{\raise.15ex\hbox{${\scriptscriptstyle -}$}\kern-.05em 1}}

\def\Dsl{\,\raise.15ex\hbox{/}\mkern-13.5mu D} 
\def\dsl{\raise.15ex\hbox{/}\kern-.57em\partial}

\font\bigit=cmti10 scaled \magstep1
\def\lspace{\ifx\answ\bigans{}\else\qquad\fi}
\def\lbspace{\ifx\answ\bigans{}\else\hskip-.2in\fi} 
\def\boxeqn#1{\vcenter{\vbox{\hrule\hbox{\vrule\kern3pt\vbox{\kern3pt
	\hbox{${\displaystyle #1}$}\kern3pt}\kern3pt\vrule}\hrule}}}
\def\mbox#1#2{\vcenter{\hrule \hbox{\vrule height#2in
		\kern#1in \vrule} \hrule}}  
%

\def\darr#1{\raise1.5ex\hbox{$\leftrightarrow$}\mkern-16.5mu #1}

\def\roughly#1{\raise.3ex\hbox{$#1$\kern-.75em\lower1ex\hbox{$\sim$}}}
\def\ss{\scriptstyle}
\def\sss{\scriptscriptstyle} \def\G{\Gamma} \def\GF{G_{\sss F}}
\def\bb{\beta\beta} \def\bbtn{\bb_{2\nu}} \def\bbm{\bb_{M}}
\def\ssbbm{\ss\bb_{M}} 
 \def\bbzn{\bb_{0\nu}}  
\def\etal{{\it et al.}}  \def\L{\Lambda} \def\phid{\phi^\dagger}
\def\phitd{\tilde\phi^\dagger} \def\phit{\tilde\phi} \def\Pr{(1+\gamma_5)}
 \def\M{{\cal M}} \def\Mn{{\cal M}_{\rm new}}
 \def\VEV#1{\left\langle #1\right\rangle}
\def\sVEV#1{\ss\langle #1\rangle}

\def\gtwid{\mathrel{\raise.3ex\hbox{$>$\kern-.75em\lower1ex\hbox{$\sim$}}}}
\def\ltwid{\mathrel{\raise.3ex\hbox{$<$\kern-.75em\lower1ex\hbox{$\sim$}}}}
 \def\frac#1#2{{\scriptstyle{#1 \over #2}}}
\def\slp{{\raise.15ex\hbox{$/$}\kern-.57em\hbox{$\partial$}}} \def\gv{g_{\sss
V}} \def\ga{g_{\sss A}}  
\def\psl{{\raise.15ex\hbox{$/$}\kern-.57em\hbox{$p$}}}
\def\qsl{{\raise.15ex\hbox{$/$}\kern-.57em\hbox{$q$}}} 
\def\Nd{$^{150}$Nd} \def\Mo{$^{100}$Mo} \def\Se{$^{82}$Se} \def\Nd{$^{150}$Nd}
\def\U{$^{238}$U} \def\Ge{$^{76}$Ge} \def\Te#1{$^{#1}$Te} \def\hbr{\hat{\bf r}}
\def\sm{{\bf\sigma}_m} \def\sn{{\bf\sigma}_n} \def\Dm{{\bf D}_m} \def\Dn{{\bf
D}_n}  \def\SN#1#2{#1\times 10^{#2}}
\def\SNt#1#2{$#1\!\times\!10^{#2}$} \def\pr#1{Phys.~Rev.~{\bf #1}}
\def\np#1{Nucl.~Phys.~{\bf #1}} \def\pl#1{Phys.~Lett.~{\bf #1}}
\def\prl#1{Phys.~Rev.~Lett.~{\bf #1}}     
\def\geff{g_{\rm eff}} \def\meff{m_{\rm eff}}

\nopagenumbers
\tolerance=10000
\hfuzz=5pt
\overfullrule=0pt
\rightline{\vbox{\hbox{McGill/92-27}
\hbox{hep-ph/9207207}
\hbox{revised version, August 1992}}}\line{}
\baselineskip=16pt plus 2pt minus 1pt
\centerline{{\bf MAJORONS FROM DOUBLE BETA DECAY$^*$\footnote{}{Talk given at
Beyond the Standard Model III, Ottawa, 1992}}}
\centerline{C.P.~BURGESS and JAMES M.~CLINE$^\dagger$\footnote{}{$\dagger$
speaker}}
\centerline{\it Physics Department,
McGill University, 3600 University Street}
\centerline{\it Montr\'eal, Qu\'ebec, Canada H3A 2T8}
\vskip 0.3 in
\centerline{ABSTRACT}
\vskip 0.3 in
\abstractfont
\baselineskip=10pt
{\narrower\narrower
We explain the existence of excess events near the endpoints of the double beta
decay ($\ss\bb$) spectra of several elements, using the neutrinoless emission
of massless Goldstone bosons.  Models with scalars carrying lepton number $-2$
are proposed for this purpose so that ordinary neutrinoless $\ss\bb$ is
forbidden, and we can raise the scale of global symmetry breaking above the 10
keV scale needed for observable emission of conventional Majorons in $\ss\bb$.
The electron spectrum has a different shape, and the rate depends on different
nuclear matrix elements, than for the emission of ordinary Majorons.\par}
\tenpoint
\vskip 0.3in

Double beta decay ($\bb$) is an extremely rare process in which two neutrons in
a nucleus simultaneously decay into protons.  It has now been observed in seven
elements: \Ge, \Se, \Mo, \Te{128}, \Te{130}, \Nd\ and \U. One reason $\bb$ has
received so much attention is the possibility of detecting small Majorana
masses for the neutrinos if the neutrinoless mode $\bbzn$ should be observed.
The amplitude for $\bbzn$ is proportional to an effective Majorana mass,
$m_{\rm eff} = \sum \theta_i^2 m_i$, where $\theta_i$ is the mixing angle
between $\nu_e$ and the $i$th mass eigenstate.  Experimentally it is known that
	\eqn\masslimit{m_{\rm eff} < 1 {\rm \ eV}.}
It is also possible that $\bbzn$ proceeds by the annihiliation of the virtual
neutrinos,\ref\Doi{M.~Doi, T.~Kotani and E.~Takasugi, \pr{D37} (1988)
2572.}$^,$\ref\georgi{H.M.~Georgi, S.L.~Glashow and S.~Nussinov, \np{B193}
(1981) 297.}\ through a vertex
	\eqn\interaction{g\varphi\bar\nu_e\gamma_5\nu_e,}
where a scalar particle $\varphi$ is emitted.  This can naturally occur if
lepton number is broken spontaneously, so that neutrinos couple to a massless
Goldstone boson called the Majoron.  The sum-energy electron spectrum for this
process, which we refer to as $\bbm$, is skewed toward higher electron
energies than that for $\bbtn$, because it is a three-body decay.  Previous
experiments searching for this mode obtained a limit\ref\nomajoron{P.~Fisher
\etal, \pl{B192} (1987) 460; D.O.~Caldwell \etal, \prl{59} (1987) 1649.}\ of
$g<\SN{2}{-4}$, following indications of a positive
signal\ref\Avignone{F.T.~Avignone III \etal, in {\it Neutrino Masses and
Neutrino Astrophysics,} proceedings of the IV Telemark Conference, Ashland,
Wisconsin, 1987, edited by V.~Barger, F.~Halzen, M.~Marshak and K.~Olive (World
Scientific, Sinagpore, 1987), p.~248.}\ at the $\SN{3}{-4}$ level.

It is therefore intriguing that the UC Irvine group has for the last few years
been seeing excess events near the endpoint in three different elements: \Se,
\Mo\ and \Nd.\ref\Moe{M.K.~Moe, M.A.~Nelson, M.A.~Vient and S.R.~Elliott,
preprint UCI-NEUTRINO 92-1 (1992).}\  A similar excess is also plainly visible
in the \Ge\ spectrum of Avignone \etal\ref\ge{F.T.~Avignone III
\etal, \pl{B256} (1991) 559.}\  In all four elements, the excess events become
evident starting from $0.5$ to 1 MeV  below the endpoint, and constitute
between 2 and 3 \% of the total signal. Because the expected signal from
$\bbtn$ is negligible near the endpoint, the chance of these events being due
to
statistical fluctuations is only 1 in $10^5$.

Let us write the partial rate for $\bbm$ as
	\eqn\rate{d\Gamma_{0\nu\sss M} = {2\pi^{-5}}
	(\ga\GF)^4 \geff^2 |\M|^2 d\Omega,}
where $\geff$ is a model-dependent effective coupling, $\M$ is the usual
combination of Gamow-Teller and Fermi nuclear matrix elements, and $d\Omega$
is the phase space.  We have analyzed the data by matching the number of excess
events above some threshold energy where visually the anomalies seem to begin
for each element. For \U\ and the \Te{}\ isotopes, only the total rate is
observed, so we omit thresholds for these elements to see how large a Majoron
coupling would be needed for $\bbm$ to saturate the total observed signal.

The results are shown in Table 1.  We find that the required values of $\geff$
are in the range $\SN{5}{-5}$ to $\SN{2}{-4}$ except for \Mo, which needs a
somewhat larger coupling.  This presents a problem for the ratio of the rates
of \Te{130}\ to \Te{128}\ decays, which has recently been
measured\ref\Haxton{W.~Haxton, quoting experimental results of Bernatowicz and
Holenberg, Neutrino `92, Granada, Spain.  It was Haxton who first used the
\Te{}\ decays  in order to constrain Majoron couplings.}\ to be $(2.41\pm
0.06)\times 10^3$.  The $\bbm$ prediction implied by Table 1 (and the phase
space calculations not shown there) would be much smaller, 93, assuming $\geff
= \SN{1}{-4}$ so that the endpoint anomalies could be explained.

\midinsert
{\noindent \ninerm Table 1: couplings for emission of ordinary Majorons in
double beta decay, assuming total rate of geochemically observed decays was
saturated by $\ss\bbm$, and using the nuclear matrix elements of
ref.~\tref\Staudt{A.~Staudt, K.~Muto and H.V.~Klapdor-Kleingrothaus,
Europhys.~Lett., {\bf 13} (1990) 31.}.}\vskip -16pt
$$\vbox{
\tabskip=0pt \offinterlineskip
\halign to \hsize{\strut#& \vrule#\tabskip 1em plus 2em minus .5em&
\hfil#\hfil &\vrule#& \hfil#\hfil &\vrule#& \hfil#\hfil &\vrule#&
\hfil#\hfil &\vrule#& \hfil#\hfil &\vrule#& \hfil#\hfil &\vrule#&
\hfil#\hfil &\vrule#& \hfil#\hfil &\vrule#\tabskip=0pt\cr
\noalign{\hrule}
&& Element && \Ge\ && \Se\ && \Mo\ && \Nd\ && \Te{128}\ && \Te{130}\ && \U\
&\cr
\noalign{\hrule}
&& $\geff$ && \SNt{1}{-4} && \SNt{8}{-5} && \SNt{4}{-4} && \SNt{8}{-5}
&&\SNt{4}{-5} &&\SNt{1}{-4} && \SNt{2}{-4} & \cr
\noalign{\hrule}
}}$$\tenpoint
\endinsert

Even if we ignore the \Te{}\ problem, there is a more serious difficulty: no
Majoron models exist which seem able to give such a large coupling.  The
favored theory for $\bbm$ was the triplet Majoron
model,$^{\georgi,}$\ref\GR{G.B.~Gelmini and M.~Roncadelli, \pl{99B} (1981)
411.}\ but it has been ruled out by LEP's measurement of the invisible $Z$
width. Alternatively, the singlet Majoron model\ref\CMP{Y.~Chikashige,
R.N.~Mohapatra and R.D.~Peccei, \pl{99B} (1981) 411.}\ has couplings which are
highly suppressed, since generically
	\eqn\suppression{g_{\rm eff} \sim m_{\rm eff} / v,}
where $v$ is the scale of lepton number breaking.  The most natural choice for
$v$ is of order the weak scale, leading to $g_{\rm eff} \sim 10^{-11}$.  But we
need $v<10$ keV to explain the size of the $\bb$ anomalies, given the limit
\masslimit\ on $m_\nu$.  This unnaturally small value requires extreme fine
tuning to keep it so far below the weak scale.\footnote{*}{{\ninerm
\baselineskip=8pt Moreover, even if  $v\sim 10$ keV, one must introduce some
neutrinos heavier than the 100 MeV scale of the nucleon Fermi momentum;
otherwise the mixing of the light neutrinos is ineffective and $\geff$ becomes
the {\it bare} coupling of the Majoron to $\nu_e$, which is zero for singlet
Majorons.\par\vskip-16pt}}  However, suppose we could remove the relation
between $v$ and $\meff$ by maintaining lepton number as an exact symmetry, thus
insuring $\meff=0$ to all orders.  Perhaps we would be able to have a higher
scale of breaking for whatever symmetry it is that gives us the Goldstone
bosons.  But if lepton number is conserved, then the Majorons must carry $-2$
units of $L$.  We will call these {\it charged Majorons} since they carry the
global lepton charge.   \pageno=2

This situation reminds us of the standard model, where a would-be Goldstone
boson, the longitudinal component of $W^-$, has electric charge.
It is therefore quite easy to construct a model of charged Majorons; we
just mimic the standard model by introducing an SU(2)$_s\times$U(1)$_{L'}$
global symmetry that acts on the leptons and on some new sterile neutrinos and
Higgs multiplets.  Although we could give a renormalizable model of this
sort,\footnote{*}{{\ninerm\baselineskip=9pt For example, introduce
sterile neutrinos $N_1$ and $N_2$ having U(1)$_{L'}$ charges $+1$ and $-1$
respectively, and include all their allowed couplings to $L_e$, $H$, $\phi$ and
$s$. Integrating out $N_i$ results in the effective theory.  Even if $N_i$ is
no heavier than the remaining particles, the resulting prediction is
quantitatively similar to that of the renormalizable theory.\par}}
it is clearest to present an effective Lagrangian with the dimension-five
operators
	\eqn\lagrangian{{g\over\L}(\overline L_e H)\Pr(\phid s)
	+ {h\over\L} (\phid \bar s)\Pr(\phitd s)+{\rm h.c.},}
where $\phit\equiv \tau_2 \phi^*$.  These particles transform as
$L_e\sim(1,1,2,
-1/2)$, $s\sim(2,0,1,0)$, $\phi\sim(2,1,1,0)$ and $H\sim(1,1,2,-1/2)$ under the
SU(2)$_s\times$U(1)$_{L'}\times$SU(2)$_L\times$U(1)$_y$ symmetries.
When $\phi$ gets a VEV $v$, a Dirac mass term between $s_1$ and
$s_2$ results. $\nu_e$ is mostly a massless state $\nu$, but mixes with a
heavy Dirac neutrino $\Psi$ with mixing angle $\theta =
\tan^{-1}(g\VEV{H}/hv)$.
One finds that the charged Majorons couple to the neutrinos via
   \eqn\coupling {{1\over 4v}\left(\sin\theta\bar\nu
	-\cos\theta\overline\Psi\right)(1-\gamma_5)
	\slp\varphi \Psi^c	+{\rm h.c.}.}
They have charge $-2$ under the residual lepton symmetry which is given
by $L=L'$ plus the diagonal generator of SU(2)$_s$.

We note that the sum-energy electron spectrum for emission of charged Majorons
differs from that of ordinary Majorons.  Consider singlet Majorons: in the
language where they are derivatively coupled ($v^{-1}\bar\nu\slp\varphi\gamma_5
\nu$), there exist extra Feynman diagrams with Majoron Brehmsstrahlung off the
{\it electron} lines via the vector coupling $v^{-1}\bar e\slp\varphi e$, while
the neutrinos annihilate through their Majoron mass. These survive at zero
Majoron momentum $q$ because the internal electrons are going on shell as
$q\to0$.  But these graphs don't exist for charged Majorons because there are
no Majorona masses; hence the amplitude vanishes as $q\to 0$, and consequently
the sum energy spectrum of the electrons has a shape intermediate between that
of $\bbtn$ and generic $\bbm$ spectra.  Our spectrum behaves like $(Q-E)^3$
near the endpoint energy $Q$,  compared with $(Q-E)^5$ for $\bbtn$ and $(Q-E)$
for ordinary $\bbm$.   The different spectra are shown in Fig.~1.

\midinsert
\vskip 2.5in
{\ninerm Fig.~1: Two-neutrino and Majoron emitting electron sum energy spectra
for\Ge.} \endinsert

Furthermore $\bbm$ with charged Majorons depends on different nuclear matrix
elements than the usual ones, because the amplitude contains an extra
factor of the neutrino momentum $p$.  The leptonic part has the Lorentz
structure $\bar e_{\sss L}\gamma_\mu[\psl,\qsl]\gamma_\nu e_{\sss L}^c$, and
for kinematic reasons only the spatial components of $p$ contribute
significantly.  Since they have odd parity, but the nuclear transition is
$0^+\to 0^+$, we require nuclear operators with odd parity also, unlike the
Gamow-Teller and Fermi terms.  Such operators come from the recoil corrections
to the hadronic weak currents, as well as p-wave Coulomb corrections to the
electron wave functions.  For example, we have matrix
elements of the form
	\eqn\ournme{{\cal M}_{\sss R} = (\gv/\ga)\VEV{p^2}^{-1}\VEV{\hbr\cdot
	(\Dn\times\sm+\sn\times\Dm) {\partial\over\partial r}h(r)},}
using the notation of Doi \etal\ref\reviewD{M.~Doi, T.~Kotani and
E.~Takasugi, Prog.~Theo~Phys.~Suppl.~{\bf 83} (1985) 1.}\ or
Tomoda.\ref\reviewT{T.~Tomoda, Rept.~Prog.~Phys.~{\bf 54} (1991) 53.}\

We find our process is roughly comparable to the ordinary Majoron process with
a coupling of
	\eqn\geffii{\geff \sim (M/2v)\theta^2
	{\VEV{qp}M^2\over(\VEV{p^2} + M^2)^2}
	{{\cal M}_{\rm new}\over {\cal M}},}
where $M$ is the mass of the heavy sterile neutrino, $\theta$ is its mixing
angle to $\nu_e$, and $\VEV{qp}$ is of order 100 MeV$^2$ since the Majoron
only has energy $q\sim 1$ MeV, whereas the virtual neutrino momentum is $\sim
100$ MeV.  ${\cal M}_{\rm new}$ denotes the new nuclear matrix elements
associated with charged Majorons. Note that $M\sim v$ is also the scale
at which our global symmetry breaks.  To get $\geff$ as large as possible we
need $\theta\sim 0.1$, which forces us to take $M > 500$ MeV due to the
constraints from peak searches in meson decays.\ref\Kconstraints{R.E.~Shrock,
\pr{D24} (1981) 1232; T.~Yamazaki \etal, {\it Proceedings of the XIth
International Conference on Neutrino Physics and Astrophysics,} eds.
K.~Kleinknecht and E.A.~Paschos (World Scientific, Singapore, 1984),
p.~183.}$^,$\ref\piconstraints{D.I.~Britton \etal, \prl{68} (1992) 3000.}\
With these values, we would also need the new nuclear matrix elements (some of
which have not been computed, to our knowledge) to be 25 times larger than the
usual ones.  Another possibility is that the coupling producing the heavy
neutrino mass is very strong ($M>v$), and that it decays so quickly into light
neutrinos and Majorons that it is really a rather broad resonance.  Then its
mass might be closer to 100 MeV and yet escape notice in $\pi$ and $K$ decay
peak searches, even with a large mixing angle.

More exactly, the rate for charged Majoron emission is,
	\eqn\chargedrate{d\Gamma_{\sss M} =
	2\pi^{-5}(\ga\GF\theta)^4(M/2v)^2|\Mn|^2 d\Omega_{\rm new}.} The
detailed forms of $\Mn$ and $d\Omega_{\rm new}$ are given in
ref.~\tref\us{C.P.~Burgess and J.M.~Cline, McGill preprint 92-22 (1992).}.
Some of the contributions to $\Mn$, such as the one displayed in eq.~\ournme,
have been calculated.  In Table 2 we compare the value of $\Mn$ needed to
account for the anomalous events with the computed value\ref\muto{K.~Muto,
E.~Bender and H.V.~Klapdor, Z.~Phys.~{\bf A 334} (1989) 187, as tabulated in
ref.~\reviewT.}\ of ${\M}_{\sss R}$, one of the actual contributions to $\Mn$.
(For this comparison we have assumed a large mixing angle and strong coupling,
as indicated in the table caption.)  The agreement is remarkably good except
for \Te{128}. However, the absolute rate \Te{128}\ is known less well than the
ratio to \Te{130}, and our model predicts $\G$(\Te{130})$/\G$(\Te{128})$ =
770$, in much better agreement with the experimental value quoted above than in
the singlet Majoron model.

\midinsert
{\ninerm\noindent Table 2: matrix elements for charged Majoron
emission in double beta decay.  We give the values of the matrix elements which
are needed to explain the data (assuming all events were $\ssbbm$ for \U\ and
\Te{}) for $\ss\theta^2(M/2v)M^2/(\sVEV{p^2}+M^2) =3\times 10^{-2},$
and values of some representative matrix elements that have been
calculated.$^{\muto}$}
\vskip -16pt
$$\vbox{
\tabskip=0pt \offinterlineskip
\halign to \hsize{\strut#& \vrule#\tabskip 1em plus 2em minus .5em&
\hfil#\hfil &\vrule#& \hfil#\hfil &\vrule#& \hfil#\hfil &\vrule#&
\hfil#\hfil &\vrule#& \hfil#\hfil &\vrule#& \hfil#\hfil &\vrule#&
\hfil#\hfil &\vrule#& \hfil#\hfil &\vrule#\tabskip=0pt\cr
\noalign{\hrule}
&& Element && \Ge\ && \Se\ && \Mo\ && \Nd\ && \Te{128}\ && \Te{130}\ && \U\
&\cr
\noalign{\hrule}
&& $|\Mn|$ needed && 1.1 && 0.99 && 1.4 && 1.3 && 0.32 && 0.48 && 0.61 & \cr
\noalign{\hrule}
&& $|\M_{\sss R}|$ (Muto \etal) && 1.1 && 0.95 && 1.1 && 1.3 && 0.92 && 0.78 &&
? &\cr
\noalign{\hrule}
}}$$\endinsert

In summary, we have presented a model of Goldstone boson emission in double
beta decay which may be able to explain excess events near the endpoints of
several elements.  The boson has lepton number $L=-2$ and the light neutrinos
remain exactly massless so that ordinary neutrinoless double beta decay is
forbidden.  The electron spectrum for these double beta decays with ``charged
Majoron'' emission is less peaked toward high energies than that for ordinary
Majorons.  The model requires a sterile Dirac neutrino with a large ($\sim
0.1$) mixing angle and hence (due to experimental constraints) a mass $M\sim
500$ MeV.  Moreover, the $\bbm$ rate in this case depends on unknown nuclear
matrix elements $\Mn$, different from those appearing in the usual amplitudes
for $\bbm$, which must turn out to be larger than the usual ones in order for
the rate to be as large as is seemingly observed.

We warmly thank M.~Moe and F.~Avignone for information about their experiments,
E.~Takasugi and T.~Kotani for helpful correspondence, W.~Haxton, M.~Luty and
P.~Vogel for valuable discussions, and R.~Fernholz for producing plots of the
spectra.  This work was supported in part by the Natural Sciences and
Engineering Research Council of Canada and les Fonds F.C.A.R. du Qu\'ebec.
\bigskip
\footatend\immediate\closeout\rfile\writestoppt
\baselineskip=8pt\bigskip\leftline{{\bf References}}\bigskip{\frenchspacing%
\parindent=20pt\escapechar=` \input refs.tmp\vfill\eject}\nonfrenchspacing
\end